\documentclass[%
 reprint,
superscriptaddress,
 amsmath,amssymb,
 aps,
floatfix,
]{revtex4-2}

\usepackage{xcolor}
\usepackage{graphicx}
\usepackage{dcolumn}
\usepackage{bm}
\usepackage{float}
\usepackage{hyperref}

\begin{document}

\preprint{APS/123-QED}

\title{Building rigid networks with prestress and selective pruning}

\author{Marco A. Galvani Cunha}
\email{mgalvani@sas.upenn.edu}
\affiliation{%
 Department of Physics \& Astronomy, University of Pennsylvania\\
}%

\author{John C. Crocker}%
\affiliation{%
Department of Chemical and Biomolecular Engineering, University of Pennsylvania}

\author{Andrea J. Liu}%
\affiliation{%
 Department of Physics \& Astronomy, University of Pennsylvania\\
}

\date{\today}

\begin{abstract}

Biopolymer networks from the intracellular to tissue scale display high rigidity and tensile stress while having coordinations well below the normal threshold for mechanical rigidity.  The elastic filaments in these networks are often severed by enzymes in a tension-inhibited manner. The effects of such pruning on the mechanics of prestressed networks have not been studied.  We show that networks pruned by a tension-inhibited method remain rigid at much lower coordinations than randomly pruned ones. These findings suggest a possible reason for the repeated evolution of tension-inhibited filament-severing proteins.
\end{abstract}

\maketitle



Biopolymer networks such as the actomyosin cell cortex, collagen extracellular matrix and fibrin blood clots play the critical role of maintaining rigidity in cells and tissues under large and often dynamically changing mechanical stresses. These living networks are remarkably stiff; the actomyosin cortex~\cite{solon_fibroblast_2007,mizuno_nonequilibrium_2007,hoffman_cell_2009,tee_cell_2011,vargas-pinto_effect_2013,rigato_high-frequency_2017} is far stiffer than one would expect from existing models of stressed networks~\cite{arzashShearinducedPhaseTransition2021,arzashFiniteSizeEffects2020,sharmaStraincontrolledCriticalityGoverns2016,licupStressControlsMechanics2015,boseSelfstressesControlStiffness2019,alexanderAmorphousSolidsTheir1998,damavandiEnergeticRigidityUnifying2022,damavandiEnergeticRigidityII2022,merkelMinimallengthApproachUnifies2019,cuiTheoryElasticConstants2019}. A key difference between the living and model networks is that the living ones exist in a dynamic steady state, with filament `edges' between nodes, as well as the nodes themselves, constantly being added and removed with the help of many proteins~\cite{fritzscheAnalysisTurnoverDynamics2013,mukhinaAActininRequiredTightly2007,wilsonMyosinIIContributes2010}. Most notably, there are  proteases (cofilin for actin\cite{mccallCofilinDrivesRapid2019,galkinActinFilamentsTension2012, pavlovActinFilamentSevering2007,schrammActinFilamentStrain2017,hayakawaActinFilamentsFunction2011a}, collagenase for collagen\cite{nabeshimaUniaxialTensionInhibits1996, yi_mechanical_2016, saini_tension_2020} and plasmin for fibrin\cite{cone_inherent_2020}) that preferentially sever and depolymerize low-tension filaments. These proteases have evolved independently in actin, collagen and fibrin networks, respectively, suggesting that they give the networks mechanical properties that confer important advantages. 

Severing (pruning) confers on each edge a binary adaptive degree of freedom, namely the possibility of being there or not there. Preferential pruning of low-tension filaments can be considered a ``local rule"~\cite{sternLearningNeuronsPhysical2023} for adjusting these degrees of freedom. Here we ask why this local rule might be so useful. 
We show that prestrained 3D central-force spring networks formed by tension-inhibited pruning are stiffer and display a narrower filament tension distribution at biologically-relevant coordinations than randomly pruned networks.  This allows the selectively pruned networks to achieve cortex-like stiffness for biophysically plausible tensile stresses and filament tensions, while comparable randomly pruned networks do not. These results help rationalize why tension-inhibited pruning has evolved independently in the three different filamentous networks.

\begin{figure}[t!]
\includegraphics[width=\linewidth]{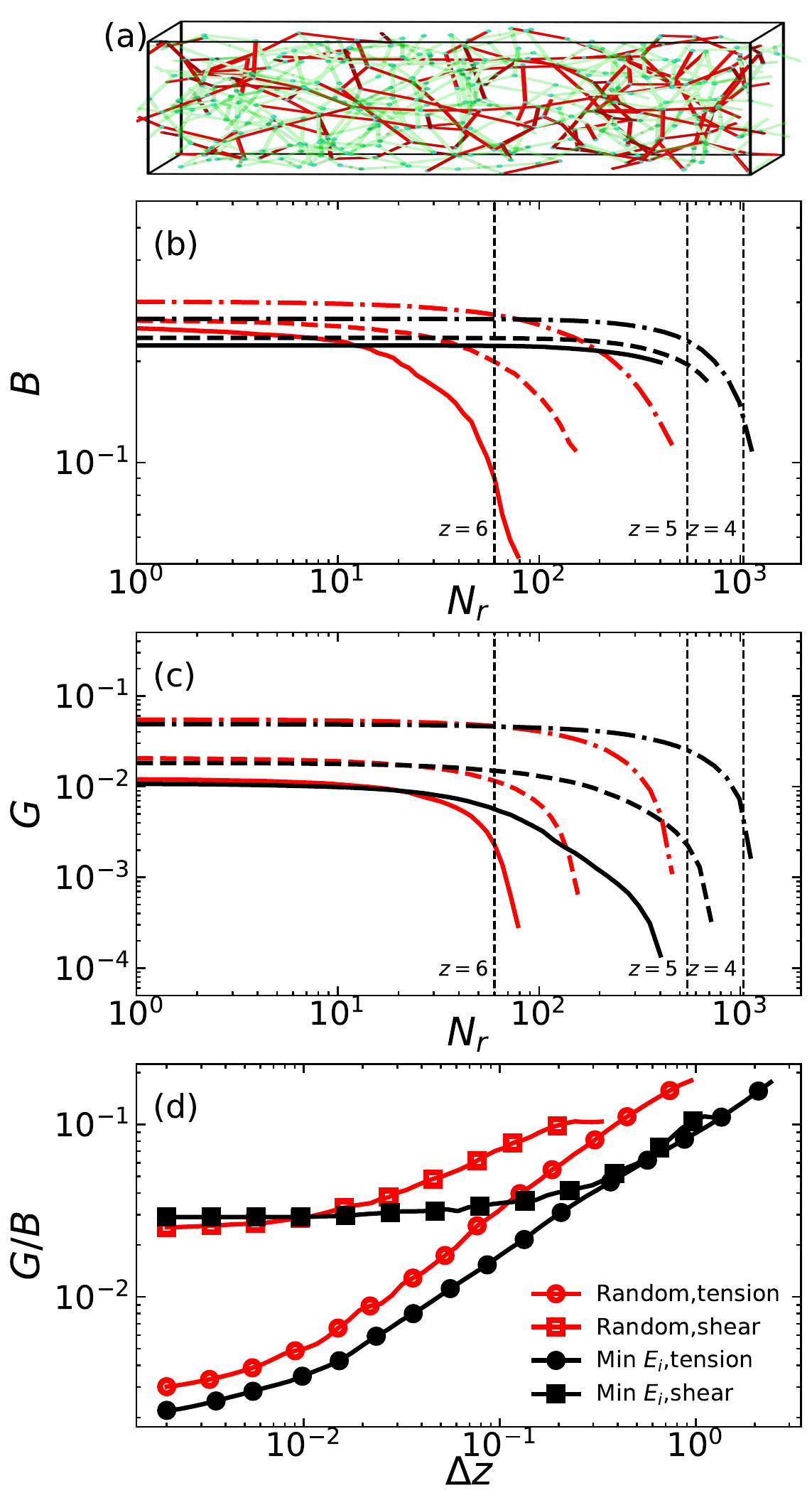}
\caption{\label{fig:BG} (a) Cross-sectional slab cut from a 3d min $E_i$ pruned network at coordination $z \approx 4$, initially under a isotropic tension prestrain of 0.05. Edges with tension higher than $3 \times 10^{-9}$ are colored red, while lower tension edges are green and transparent. The actual simulation box is larger than the slab depicted and fully periodic.
(b,c) Bulk (b) and shear (c) moduli of networks under isotropic tension during random (red) and min $E_i$ (black) pruning as a function of the number of bonds removed $N_r$. Different lines (solid, dashed, dotted) indicate different strains $\epsilon$ used to initially prestress the network (0.0001, 0.001, 0.01 respectively). (d) Ratio of shear to bulk modulus, $G/B$ vs.~$\Delta z = z-z_c$, the difference from current coordination to the critical coordination, for isotropic tension (circles) and pure shear (squares) for random (red) and min $E_i$ (black) pruning. Networks were prepared with a prestrain of 0.01.}
\end{figure}

Central-force spring networks exhibit a transition from a rigid to a floppy state as a function of their coordination $z$. In the absence of edge tension this transition happens at the isostatic value $z_{iso} = 6$ (in 3D).  Such networks can remain rigid below $z=6$, however, if they are under prestress~\cite{arzashShearinducedPhaseTransition2021,arzashFiniteSizeEffects2020,sharmaStraincontrolledCriticalityGoverns2016,licupStressControlsMechanics2015,boseSelfstressesControlStiffness2019,alexanderAmorphousSolidsTheir1998,damavandiEnergeticRigidityUnifying2022,damavandiEnergeticRigidityII2022,merkelMinimallengthApproachUnifies2019,cuiTheoryElasticConstants2019}. We generate our networks from jammed packings of bidisperse particles with initial $z>6$ by placing $N = 1024$ soft repulsive particles at random in a three-dimensional box at a volume fraction above jamming and minimizing the total energy. 
We then use the geometry of the jammed state to generate unstressed networks by placing nodes at the centers of particles and relaxed springs, with rest length equal to the spring length, between nodes. We use periodic boundary conditions throughout the simulations. We next apply an initial prestrain by changing the rest lengths of all of the springs in one of three ways: isotropic tension, pure shear or Gaussian random strains. Changes in rest lengths $r_0$ are given by:
\begin{equation}
    r_0 \rightarrow \begin{cases}
    (1-\epsilon)r_0 & \text{isotropic tension} \\
    (1-\epsilon[\hat r_{y}^2-\hat r_{x}^2])r_0 & \text{pure shear} \\
    (1-\eta)r_0 & \text{Gaussian random}
    \end{cases}
\end{equation}
\noindent where $\epsilon \ll 1$ is a small prestrain, $\hat r_{i}$ is the i-th component of the unit vector that points along the edge, and $\eta$ is a Gaussian random variable with zero mean and standard deviation $\epsilon$. In the isotropic tension case, all edges are tensed.  In the pure shear and random cases, some edges are under compression while some are under tension. While the simulation box is kept fixed during our simulations, the edge strains induced by the change in the rests lengths are equal to lowest order to the strains induced by an equivalent deformation of the simulation box for the isotropic and pure shear cases. The Gaussian random prestress cannot be directly mapped to a box strain. After changing the rest lengths the nodes are not at rest, so we use FIRE \cite{guenoleAssessmentOptimizationFast2020} to minimize the network to a force-balanced state and obtain the energy $E_i$ stored in each edge. 

We study the effects on these networks of sequential pruning according to two different protocols: random pruning and pruning only the edge with the lowest $E_i$, ``\textit{min $E_i$}" pruning. 
To avoid generating 2-fold coordinated nodes we prune only edges connected to nodes that have coordinations $\ge 4$. It is interesting to note that strategies based on pruning edges with \emph{maximal} $E_i$ lead to fracture of the networks for all types of prestrain. 

After each pruning event we minimize the energy to maintain a force-balanced state. From this state, we calculate the moduli using linear response. We prune until we reach the critical coordination $z_c$, defined as where the bulk modulus drops by more than an order of magnitude within one pruning event. The linear response calculation is an extension to nonzero prestress of a previous approach\cite{goodrichPrincipleIndependentBondLevel2015}. Essentially, we apply an affine deformation to the network and calculate the induced non-affine displacements due to the affine forces by using the Hessian. The resulting total displacement allows us to calculate the change in energy at each edge for a given deformation, which gives the contribution of that edge to the corresponding modulus (see Supplemental Material, SM). A representative cross-section cut from a rigid network at $z \approx 4$ is shown in Fig.~\ref{fig:BG}(a).  

It has been shown that pruning edges in \emph{unstressed} central-force spring networks with mean coordinations $z>6$ leads to mechanical properties that are extremely sensitive to choice of which bonds are pruned~\cite{goodrichPrincipleIndependentBondLevel2015,hexnerLinkingMicroscopicMacroscopic2018,hexnerRoleLocalResponse2018}. Fig. \ref{fig:BG}(b,c) shows the bulk and shear moduli $B$ and $G$ as a function of the number of edges removed, $N_r$, during random or min $E_i$ pruning of a network initially placed under isotropic tension. Curves are shown for networks prestrained with $\epsilon$ ranging from 0.0001 to 0.01, with vertical lines showing the corresponding coordination of the networks at a given $N_r$. Recall that pruning terminates when the network becomes floppy at $z_c$. We find that the behavior of $G/B$ (Fig.~\ref{fig:BG}(d)) as $z \rightarrow z_c^+$ does not depend on pruning protocol; this is because the contributions of each bond $i$ to the bulk and shear moduli, $B_i$ and $G_i$, are strongly correlated by the prestress (see SM). 

In overcoordinated central-force spring networks placed under a small isotropic tension,  we would expect $G/B \rightarrow 0$  for min $E_i$ pruning and $G/B \rightarrow \mathrm{const}$ for random pruning as $z \rightarrow z_c$~\cite{goodrichPrincipleIndependentBondLevel2015,hexnerLinkingMicroscopicMacroscopic2018}. In contrast, in under-coordinated networks under a significant isotropic tension we find $G/B \rightarrow 0$ for \emph{both} pruning protocols (Fig.~\ref{fig:BG}(d)). (The upturn at the lowest values of $\Delta z$ in Fig.~\ref{fig:BG}(d) is a finite-size effect (see SM)). The result $G/B \rightarrow 0$ is consistent with earlier findings that $G \rightarrow 0$ but $B \rightarrow \mathrm{const}$ as $z \rightarrow z_c^+$ for under-coordinated networks under isotropic tension\cite{merkelMinimallengthApproachUnifies2019,arzashStressstabilizedSubisostaticFiber2019,arzashMechanicsFiberNetworks2022,sheinmanActivelyStressedMarginal2012}. 

For over-coordinated networks placed under a small pure shear, we would expect $G/B \rightarrow \infty$ for min-$E_i$ pruning but $G/B \rightarrow \mathrm{const}$ for random pruning, as $z \rightarrow z_c$ ~\cite{goodrichPrincipleIndependentBondLevel2015,hexnerLinkingMicroscopicMacroscopic2018}. By contrast, for under-coordinated networks under significant pure shear we find $G/B \rightarrow \mathrm{const}$ for both pruning protocols (Fig.~\ref{fig:BG}(d)). This is consistent with earlier findings that $G$ and $B$ both approach constants as $z \rightarrow z_c^+$ for such systems under pure shear~\cite{merkelMinimallengthApproachUnifies2019} (see SM); such behavior is also observed in shear-jammed packings~\cite{baityjesiEmergentSO32017}.  

\begin{figure}[t!]
\includegraphics[width=\linewidth]{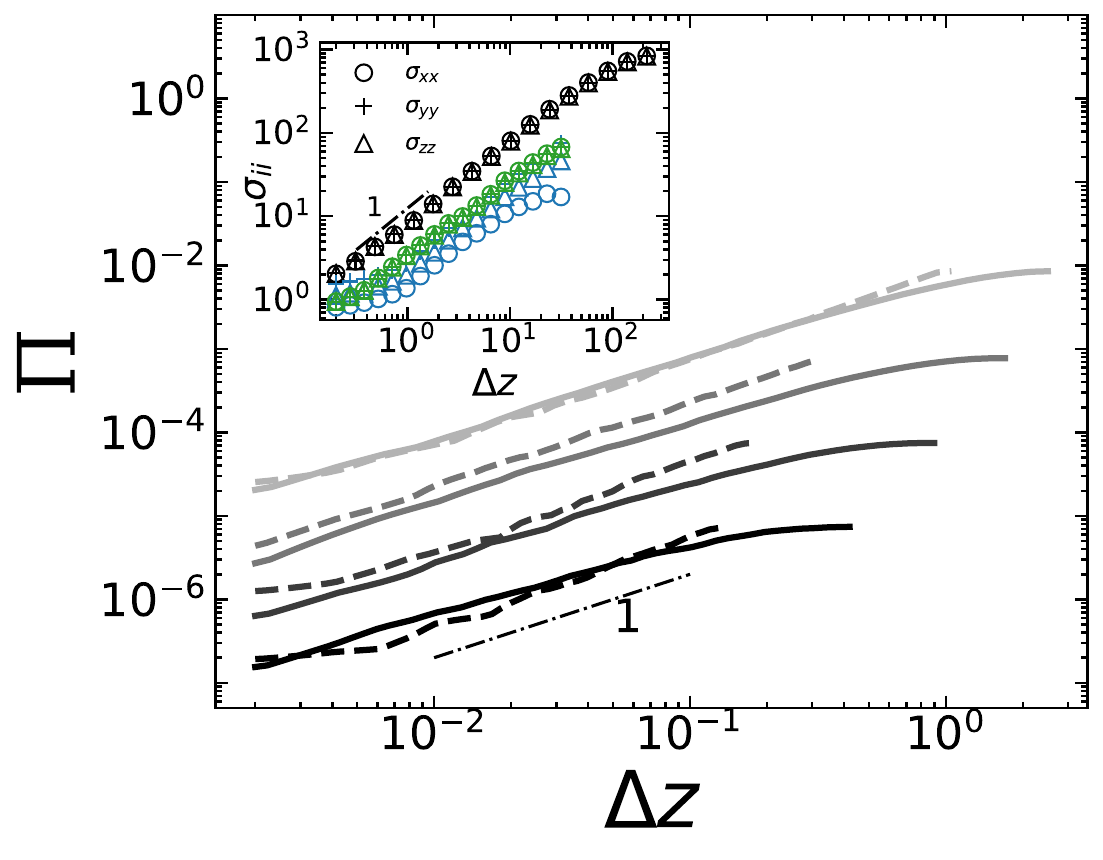}
\caption{\label{fig:stress} Tension on randomly (dashed) or min $E_i$ (solid) pruned isotropic tension networks as a function of 
$\Delta z$. The different shades represent different initial prestrains ($0.00001$, $0.0001$, $0.001$ and $0.01$ from darkest to lightest). (inset)  Diagonal components of stress for randomly pruned networks at $0.001$ initial strain for the three different types of prestrain (black for tension, blue for shear and green for Gaussian). Different symbols correspond to different diagonal components of the stress. All networks converge to a state where the three diagonal components are positive even when compressed edges are present in the initial state.} 
\end{figure}

Since the networks are floppy when unstressed, the total tension sets the scale of the bulk and shear moduli. 
Different pruning methods lead to different values of $z_c$ (Fig.~\ref{fig:BG}). It is possible that they might also lead to different \emph{scalings} for the tension as $z \rightarrow z_c^+$. Fig. \ref{fig:stress} shows the total tensile stress or `tension' of the network, $\Pi = -Tr{\bm \sigma}/3$, where $\bm{\sigma} = \frac{1}{V}\sum_{ij}r_{ij} \times f_{ij}$ is the stress tensor, $r_{ij}$ is the distance between bonded nodes and $f_{ij}$ is the force stored in bond $ij$, for the case where the initial prestress has the form of isotropic tension. We plot $\Pi$ as a function of $\Delta z = z-z_c$ for random and min $E_i$ edge pruning. Close to the transition, we see that $\Pi \sim \Delta z$, independent of the pruning method. 
The inset to Fig.~\ref{fig:stress} shows that the same scaling holds for each of the individual diagonal components of the stress tensor $\sigma_{ii}$ during random pruning for the different types of prestrains. Thus, the scaling of $\Pi$ for $z \rightarrow z_c^+$ is independent of pruning method and type of initial prestrain. The pruning method then primarily affects the scaling of the bulk and shear moduli with $\Delta z$ by lowering the critical coordination $z_c$.

Note from the inset to Fig.~\ref{fig:stress} that near the transition, all diagonal components of the stress tensor are positive, indicating the network is under tension in all directions. This is surprising since $\sigma_{xx}$ is initially negative for systems initially placed under pure shear strain by expanding in the $y$-direction and compressing in the $x$ direction. To understand this, we note that spring networks are unstable under compression due to structural buckling\cite{alexanderAmorphousSolidsTheir1998}. The observation that $\sigma_{xx}$ changes sign to become positive indicates that compressed edges are either removed or put under tension by pruning.  As a result, after sufficient pruning the network eventually reverts to a state of tension for any form of the initial prestress. 

\begin{figure}[t!]
\includegraphics[width=\linewidth]{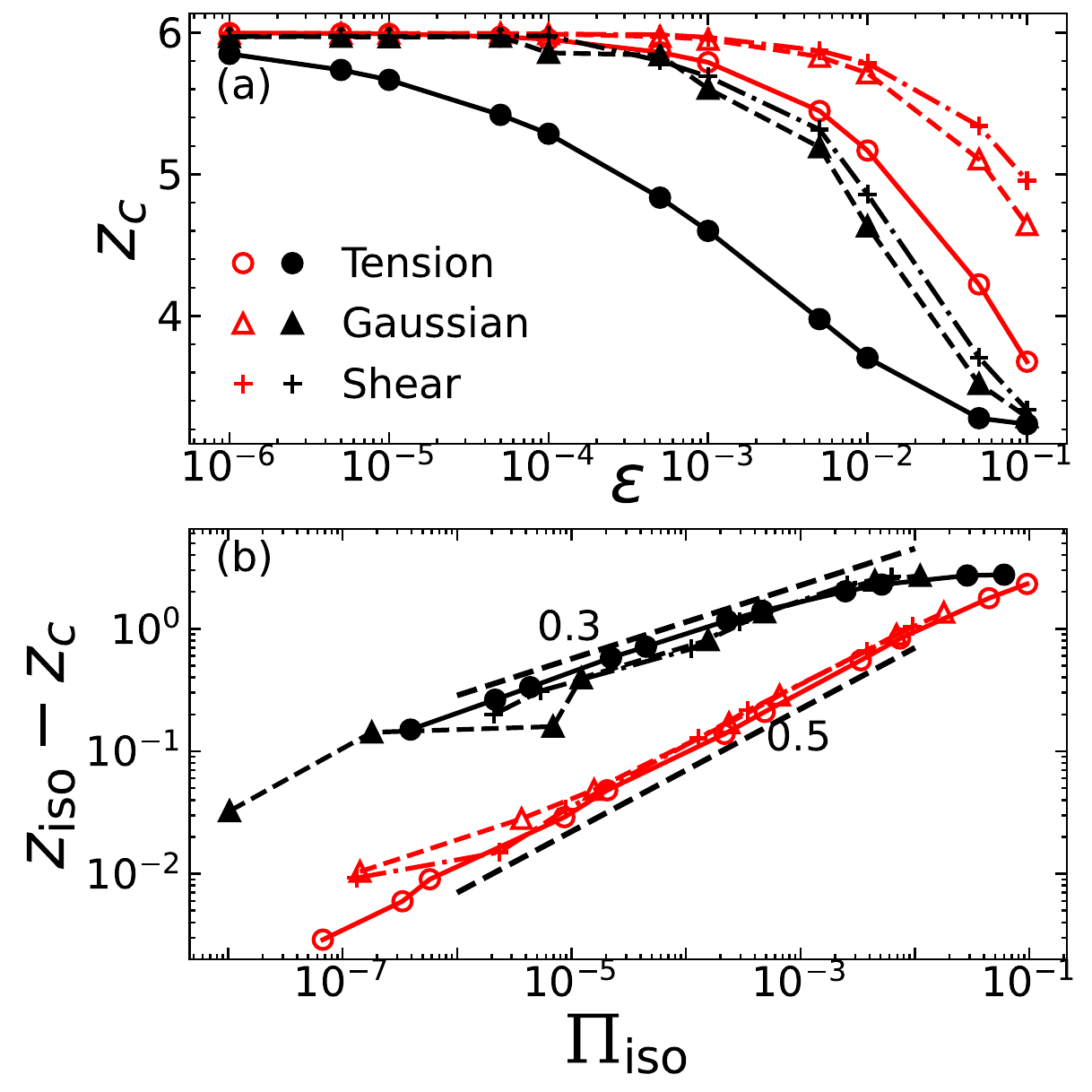}
\caption{\label{fig:scaling} (a) Critical coordination as a function of initial strain for different types of prestrain and pruning strategy. The networks under isotropic tension (solid circles) show a lower $z_c$ than networks under pure shear (dash-dot crosses) or with Gaussian distributed disorder in rest lengths (dashed triangles). Min $E_i$ pruning (black) leads to networks that are still rigid at lower coordinations compared to random pruning (red). (b) Difference between critical coordination and the isostatic value as a function of the trace of the total tension of the networks. Min $E_i$ data flattens out for the highest tension values due to the constraint on pruning 3-fold coordinated nodes. Marker and line styles and colors are the same as in (a).}
\end{figure}

We now examine the critical coordination reached by pruning in more detail. Fig. \ref{fig:scaling}(a) shows how $z_c$ depends on the initial prestrain $\epsilon$ for different types of prestrain and for the two pruning methods. In general, $z_c$ decreases with increasing $\epsilon$. This is reminiscent of the behavior of strain stabilized subisostatic networks, where the value of the critical strain at the onset of stiffness increases as a function of $\Delta z$ below the isostatic point \cite{sharmaStraincontrolledCriticalityGoverns2016,merkelMinimallengthApproachUnifies2019}. As noted earlier, $z_c$ is lower for min $E_i$-pruned networks than for randomly pruned networks at the same $\epsilon$, for all types of initial prestrain. Importantly, min $E_i$-pruned networks under isotropic tension can reach coordination values found in biopolymer networks with initial prestrains on the order of 1-5\%, while randomly-pruned networks require on the order of 10\% prestrain on the filaments to reach the same coordination.

Note that networks prepared at different initial coordinations above $z_{iso}$ require different prestrains to attain the same $z_c$ from pruning. Therefore, the initial prestrain is not very useful for characterizing a network. A more useful measure is the total amount of tension at the isostatic coordination $\Pi_{iso} = \Pi|_{z=z_{iso}}$. This is the total amount of tension available to stabilize the network at the point where the network would become floppy with no stress. In Fig. \ref{fig:scaling}(b) we plot $z_{iso}-z_c$ as a function of $\Pi_{iso}$. 

Fig.~\ref{fig:scaling}(b) shows that for all types of prestrain the results collapse onto two different curves, one for random pruning and the other for min $E_i$ pruning. For small $\Pi_{iso}$, min $E_i$ pruning leads to a decrease in $z_c$ that is an order of magnitude larger than for random pruning, and the difference between pruning methods decreases with $\Pi_{iso}$. We observe the scaling 
\begin{equation}
    z_c^{iso}-z_c \sim \Pi_{iso}^{\alpha}
    \label{eq:collapse}
\end{equation}
\noindent where $\alpha \approx 0.5$ for random pruning and $\alpha \approx 0.3$ for min $E_i$ pruning.  With this scaling, the critical coordination for a network can be predicted from $\Pi_{iso}$ for a given pruning method. Since biological filament networks have a mean coordination between 3 and 4 due to a combination of branching points and crosslinks and are rigid, model networks must have $z_c$ in that range. Fig.~\ref{fig:stress} shows that stress scales as $\Delta z$ independently of stress type or pruning method, so the isostatic stress $\Pi_{iso}$ can be used as a proxy for the stress throughout the pruning process. The smaller scaling exponent $\alpha$ means that biologically-relevant coordinations are more easily accessible by min $E_i$ pruning, requiring tensions up to an order of magnitude smaller than random pruning.

\begin{figure}[t]
\includegraphics[width=\linewidth]{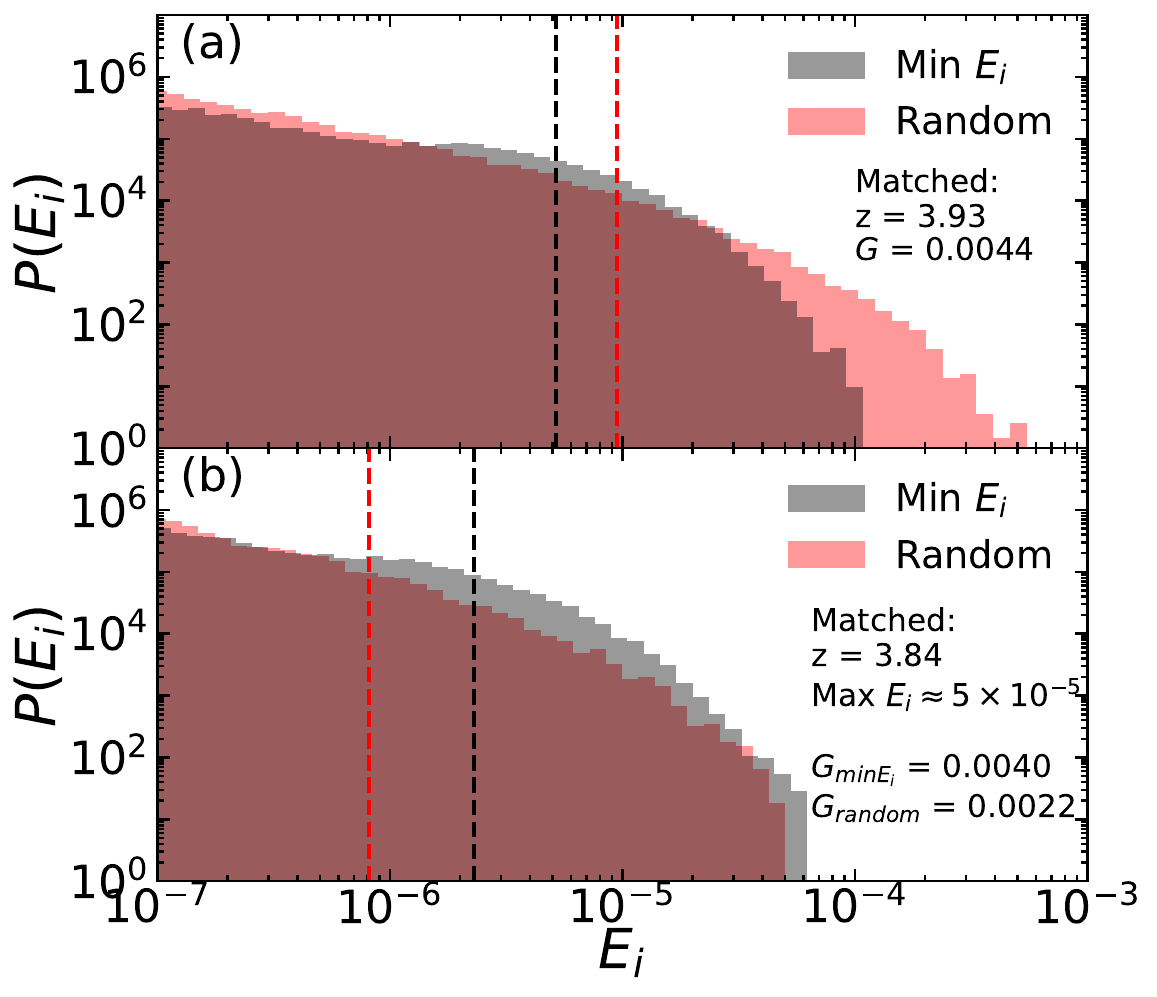}
\caption{\label{fig:BG_fixed_z} 
(a) Distribution of edge energies $E_i$ at same $z$ and $G$ for a min $E_i$ (black) and a randomly (red) pruned network. Dashed lines show the average bond energy. The randomly pruned network requires a larger average and maximum energy to sustain the same modulus as a min $E_i$ pruned network at same $z$. In our model $E_i \sim \delta_i^2$, where $\delta_i$ is the strain of edge $i$, so an energy of $10^{-4}$ corresponds to a strain of 1\%. Both networks are at tension $\Pi \approx 1.7\times10^{-3}$. (b) Distribution of edge energies as in (a), but at a lower coordination where both networks have a similar maximum $E_i$. The min $E_i$ network is at tension $\Pi \approx 1.1\times10^{-3}$, while the randomly pruned network is at a much lower tension of $\Pi \approx 1.6\times10^{-4}$.}
\end{figure}

To examine this point in more detail, we compare the properties of networks created by each pruning method at a fixed, biologically-relevant coordination. At a fixed $z$, the shear modulus obeys a power-law scaling with tension that is independent of $z$ or pruning strategy (see Fig. S4 in SM). In contrast, other biologically-relevant mechanical properties do depend on pruning strategy at fixed $z$. We compare the probability distribution of edge energies in two networks pruned by different methods at the same $z$ and shear modulus (Fig.~\ref{fig:BG_fixed_z}(a)) as well as at the same $z$ and maximum edge energy (Fig.~\ref{fig:BG_fixed_z}(b)). We accomplish this by comparing a randomly pruned network under isotropic tension prestrain of 0.08 and a min $E_i$-pruned network under a prestrain of 0.01 that have similar mechanical properties at the same $z \approx 4$.  Fig.~\ref{fig:BG_fixed_z}(a) shows that the min $E_i$-pruned network has a much narrower distribution of edge energies and also smaller mean $E_i$ when compared to the randomly-pruned network. This implies that a min $E_i$ pruned network can achieve the same shear modulus with much less average energy stored in the edges. A similar effect has been seen in networks with catch and slip bonds \cite{mullaWeakCatchBonds2022}, where catch bonds (which have smaller lifetimes at low tensions) exhibit a narrower distribution of loads per crosslinker. This suggests that narrow distributions of tension may be a universal feature of models where edges preferentially break or are removed at low tensions, and is independent of the particular failure mechanism or kinetics.
 
Fig. \ref{fig:BG_fixed_z}(b) shows that the min $E_i$ pruned network has almost double the shear modulus of a randomly pruned one when the maximum bond energies are similar. In networks where the stress is derived from the motion of the motors, stall forces function as a tension threshold for the filaments.  Alternatively, the maximum filament tension could be set by a breaking limit. Fig.~\ref{fig:BG_fixed_z}(b) shows that min $E_i$ pruning leads to stiffer networks under such filament tension-limited conditions.

\textit{Implications for actomyosin cortex.} Our results shed some light on the cortical tension $\Pi$ and apparent shear modulus $G$. Quantities in our model are expressed in terms of the network edges' spring constant $\kappa$ and mean length $l$, which can be scaled using experimental measurements.  Single actin filament stretching studies \cite{kojimaDirectMeasurementStiffness1994,liuMechanicsFActinCharacterized2002,matsushitaEvaluationExtensionalTorsional2010} show the stretching spring constant $\kappa$ of F-actin is approximately 40 pN/nm for $l = 1\mu$m, with stiffness scaling as $\kappa \sim 1/l$. 

The stiffness and tension of our model can be compared to the typical stiffness and tension of the cortex, despite significant differences in reported values of the latter amongst different methods \cite{hoffman_cell_2009}.  Careful AFM indenting measurements \cite{rigato_high-frequency_2017} combined with finite element modeling of the cortex as a $200$ nm thick viscoelastic sheet \cite{vargas-pinto_effect_2013}, suggests a value $G \approx 15$ kPa, orders of magnitude higher than for an unstressed actin gel. Micropipette aspiration of many cells report cortical tension of order 1 nN/$\mu$m \cite{kee_micropipette_2013, cartagena2016actomyosin}, corresponding to a large tensile stress of $\Pi \approx 5$ kPa for a 200 nm thick cortex. Scaling our values of the shear moduli and tensile stress in Fig.~\ref{fig:BG_fixed_z}(a), with an assumed value of $l \approx 100$ nm, yields plausible values of $G_{min E_i} \approx 4.4 \times 10^{-3} \kappa/l \approx 17.6$ kPa, and $\Pi \approx 1.7 \times 10^{-3} \kappa/l \approx 6.8$ kPa. Translating the edge energies to filament tensions $t_i \sim \sqrt E_i$ leads to an estimate for maximum tension $t_{max} = 400$ pN for min $E_i$ pruning and $t_{max} = 894$ pN for random pruning. The breaking force of actin filaments is $600$ pN \cite{tsudaTorsionalRigiditySingle1996}, between the maximal filament tensions in our two model networks.  That is, we find that our selectively pruned network model can reproduce the stiffness and tension of the actomyosin cortex with physically plausible filament tensions, while a randomly pruned network cannot.

\textit{Discussion.} It has previously been shown that for unstressed, over-coordinated networks, the scaling of the ratio of the shear to bulk modulus, $G/B$, depends sensitively on pruning protocol~\cite{goodrichPrincipleIndependentBondLevel2015,hexnerLinkingMicroscopicMacroscopic2018,hexnerRoleLocalResponse2018}. Here we have shown that pruning protocol does \emph{not} affect the scaling of the ratio of the shear to bulk modulus, $G/B$,  with $\Delta z$. This is because prestress highly correlates the changes in the bulk and shear moduli due to bond removal, in contrast to the unstressed, over-coordinated case
~\cite{hexnerRoleLocalResponse2018}. The reason why different biopolymer networks have independently evolved the same pruning strategy is therefore not to produce a desired scaling for $G/B$.

Why then are biological filament networks such as the actomyosin cell cortex, collagen extracellular matrix and fibrin blood clots, all accompanied \textit{in vivo} by proteases that impose tension-inhibited pruning? The mean coordination of these networks is significantly lower than $z_{iso}$, so significant prestress must be maintained in order for the networks to be rigid. This prestress must be supported and rigidity maintained under wildly varying extreme deformation even as some filaments are being destroyed by tension-inhibited proteins (e.g.~cofilin, collagenase, plasmin)~\cite{mccallCofilinDrivesRapid2019,galkinActinFilamentsTension2012,pavlovActinFilamentSevering2007,schrammActinFilamentStrain2017,hayakawaActinFilamentsFunction2011a, nabeshimaUniaxialTensionInhibits1996, yi_mechanical_2016, saini_tension_2020, cone_inherent_2020}. Our results show that selectively pruning low-tension edges leads to significantly smaller values of $z_c$ for the same prestrain when compared to random pruning. Selective pruning also generates networks with significantly higher shear modulus when the maximum edge tension is limited. We conjecture that the functional advantage inferred by this higher stiffness may explain the repeated evolution of the motif of tension-inhibited severing proteins  \cite{mccallCofilinDrivesRapid2019,galkinActinFilamentsTension2012,pavlovActinFilamentSevering2007,schrammActinFilamentStrain2017,hayakawaActinFilamentsFunction2011a, nabeshimaUniaxialTensionInhibits1996, yi_mechanical_2016, saini_tension_2020, cone_inherent_2020}. 

Living biological filament networks are dynamically remodeled;  new filament constantly replace severed and depolymerized ones in a homeostatic state. The living cortex undergoes large strain fluctuations with a heavy-tailed distribution~\cite{sivarajan2023evy} of amplitudes, termed cytoquakes. It seems possible that the distribution of these fluctuations may be related to the tension distribution of the most tensed filaments, as we report here, via mechanisms such as filament severing, or myosin sliding or unbinding.

In the actomyosin cortex, dynamical remodeling leads to complete turnover of actin filaments every 30 seconds or so\cite{fritzscheAnalysisTurnoverDynamics2013}, with a correspondingly large expenditure of metabolic energy. Turnover introduces an adaptive degree of freedom for each edge of the network, corresponding to whether the edge is there or not there. These adaptive degrees of freedom are adjusted by local rules, one of which is tension-inhibited pruning. Our results on tension-inhibited pruning suggest that the result of the local rules controlling turnover might be to allow the cortex to maintain the collective property of rigidity under constantly varying and complex mechanical stresses, thus `justifying' its high metabolic cost.  

We thank Sadjad Arzash, Paul Janmey, Fred MacKintosh, Ayanna Matthews, Sidney Nagel and Daniel Reich for instructive discussions. MAGC was supported by NSF through the University of Pennsylvania Materials Research Science and Engineering Center (MRSEC) since the inception of DMR-2309043, and by NSF-PHY-1915174 and NSF-DMR-2005749 prior to that. AJL thanks the Simons Foundation for support via \#327939 as well as the Center for Computational Biology at the Flatiron Institute. AJL also thanks the Isaac Newton Institute for Mathematical Sciences at Cambridge University (EPSRC grant EP/R014601/1), for support and hospitality.


\bibliography{apssamp}

\clearpage

\appendix

\section{Linear response formalism for prestressed networks}

All mechanical properties of the networks are calculated in linear response, which we adapt from previous work to include the effects of prestress. All networks considered are composed of harmonic springs with the same stretching stiffness $k_i = 1$ and no bending interactions. We calculate the change in energy of the network after a deformation given by some strain tensor $\epsilon$ from the lowest order expansion of the energy

\begin{equation}
    \Delta E = \sum_i k_i\delta r^2_{i,||}-\frac{t_i}{r_i}\delta r^2_{i,\perp}
    \label{eq:linear_energy}
\end{equation}

\noindent where \textit{i} runs over all edges of the network, $\delta r_{i,||}$ ($\delta r_{i,\perp}$) is the total strain on edge \textit{i} that is parallel (perpendicular) to the edge direction and $t_i$ is the tension on edge \textit{i} in the reference state. The second term in eq. \ref{eq:linear_energy} is called the prestress term.

The change in energy of the network can also be written as

\begin{equation}
    \frac{\Delta E}{V} = \frac{1}{2}\epsilon_{\alpha\beta}c_{\alpha\beta\gamma\delta}\epsilon_{\gamma\delta}
    \label{eq:linear_stiffness}
\end{equation}

\noindent where $c_{\alpha\beta\gamma\delta}$ is the stiffness tensor and $V$ is the volume of the network. Using both equations, we can use the edge strain induced by the deformations to calculate all the components of the stiffness tensor as well as the bulk modulus and angle-average shear modulus, given by

\begin{equation}
    \begin{split}
    B = \frac{1}{9}&(c_{xxxx}+c_{yyyy}+c_{zzzz}+2c_{yyzz}+2c_{xxyy}+2c_{xxzz})
    \\
    G = \frac{1}{15}&(3c_{yzyz}+3c_{xyxy}+3c_{xzxz}+c_{xxxx}+c_{yyyy}+c_{zzzz}\\
    &-c_{yyzz}-c_{xxyy}-c_{xxzz}).
    \end{split}
    \label{eq:moduli}
\end{equation}

To calculate the edge strains in eq. \ref{eq:linear_energy} we take a two step approach where we first calculate the affine strain induced by the strain tensor $\epsilon$ and then obtain the non-affine strain from the forces that result from the affine deformation. The non-affine strain is obtained from

\begin{equation}
    M_{\alpha\beta}\delta r^{NA}_\alpha = f_{\beta}
\end{equation}

where $\bm M$ is the Hessian matrix, which includes both unstressed and stressed components, similar to the separation observed in eq. \ref{eq:linear_energy}. The total strain is the sum of the affine and non-affine components, and can then be used to calculate the change in energy of the network under the applied strain tensor.

We also note that all mechanical properties can be broken down to their contributions from each edge in the network. Eq. \ref{eq:linear_energy} can be written as $\Delta E = \sum_i \Delta E_i$ and similar reasoning with equations \ref{eq:linear_stiffness} and \ref{eq:moduli} allows us to define the \textit{i}th edge contribution to the bulk (shear) modulus $B_i$ ($G_i$). These can be used as targets for pruning strategies along with the energy of edge \textit{i} in the reference state $E_i$, and we will show that all these quantities become correlated during pruning.

\section{Correlations between $E_i$, $B_i$ and $G_i$}

The edge contributions to the shear and bulk modulus can be used as the basis of a controlled pruning strategy and have been previously shown to lead to very efficient tuning of mechanical properties, but there is no known mechanism for cleaving proteins in biopolymer networks to measure these quantities. These proteins can however measure edge tension (which in our model is equal to the square root of $E_i$), and the difference between a tension-based pruning strategy and a $B_i$ or $G_i$ based pruning strategy will depend on the degree of correlation between $E_i$, $B_i$ and $G_i$.

We measured the Pearson correlation coefficients between $E_i$ and the edge moduli for all our networks during pruning

\begin{equation}
    corr(A,B) = \frac{\langle (A-\langle A\rangle)(B-\langle B\rangle)\rangle}{\sigma_A \sigma_B}
\end{equation}

\noindent where $\sigma_X$ is the standard deviation of $X$ and the average is taken over all edges. The resulting coefficient are shown in fig. \ref{fig:correlations}, and show that correlations are very high even in the initial states, and generally rise during pruning and with increasing prestress. For sufficiently prestressed networks that remain rigid at biologically relevant coordinations we conclude that tension-based pruning is equivalent to pruning based on local moduli, since correlations approach 1 in that regime.

\renewcommand{\thefigure}{A\arabic{figure}}

\setcounter{figure}{0}

\begin{figure*}[ht]
    \centering
    \includegraphics[width=\textwidth]{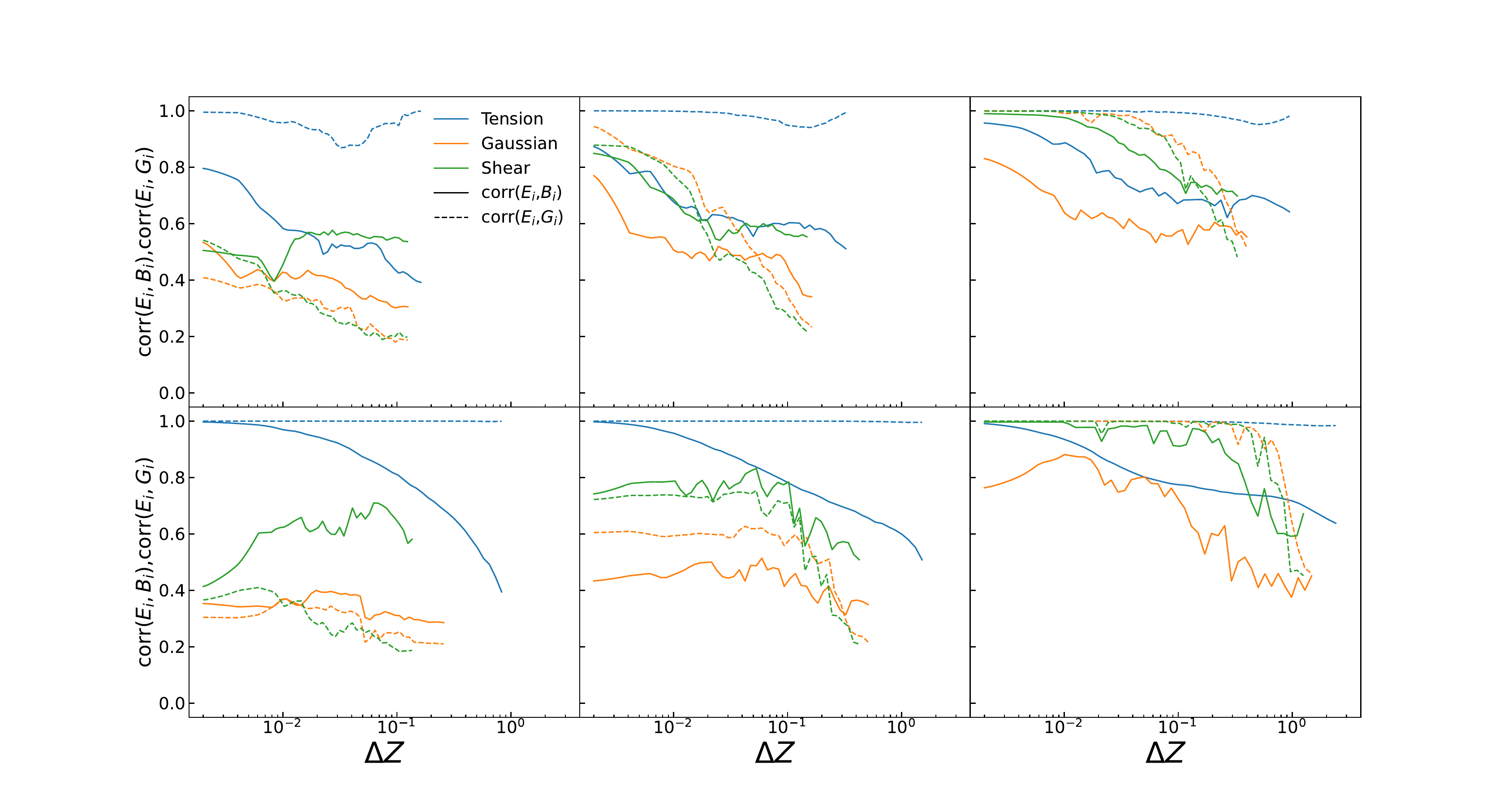}
    \caption{Pearson correlations coefficients between $E_i$ and $B_i$ (solid), and $E_i$ and $G_i$ (dashed) during random (top) and min $E_i$ (bottom) pruning. The coefficients are calculated for different types of prestress (colors, see legend on top left) and different values of initial prestrain (0.0001 (left), 0.001 (center) and 0.01 (right)).}
    \label{fig:correlations}
\end{figure*}

\section{Finite size effects on $z_c$, B and G}

\begin{figure*}[ht]
    \centering
    \includegraphics[width=0.7\linewidth]{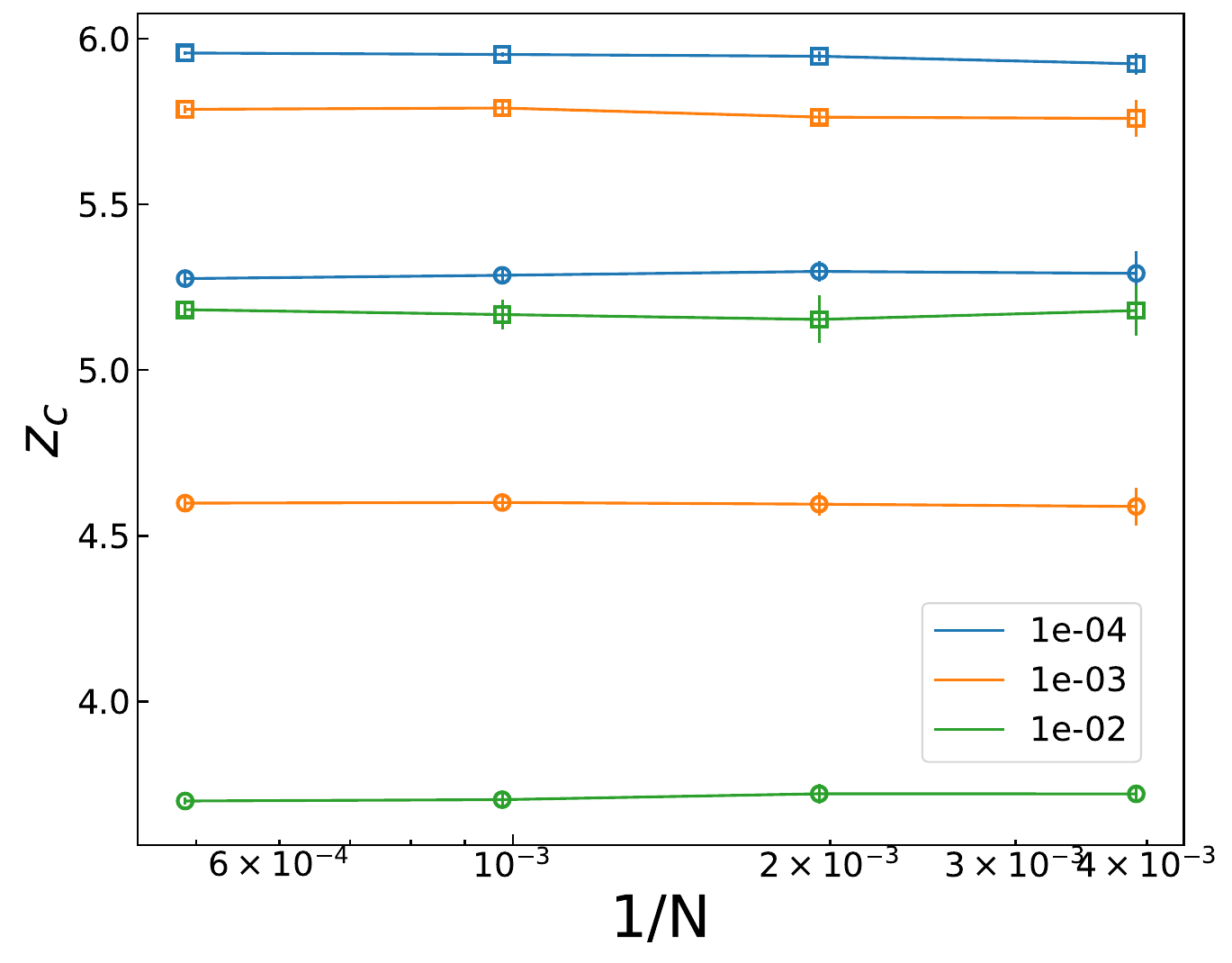}
    \caption{Critical coordination $z_c$ for networks of different number of nodes $N$ at different prestrains for both min $E_i$ (circles) and random (squares) pruning. Data points are the average of 10 simulations, with error bars representing the standard deviation. Different colors represent different nominal prestrains.}
    \label{fig:zc_finite}
\end{figure*}

In the paper, we have focused on the effect of pruning method and prestress on the value of the critical coordination $z_c$. In principle, this value could be affected by finite size effects related to the size of our networks. In order to account for this, we have performed pruning in networks with a different number of nodes $N$ ranging from 256 to 2048. Fig. \ref{fig:zc_finite} shows that the critical coordination is independent of the network size, suggesting that the values of $z_c$ reported in the main manuscript are accurate despite the size of the network.

\begin{figure*}[ht]
    \centering
    \includegraphics[width=0.8\textwidth]{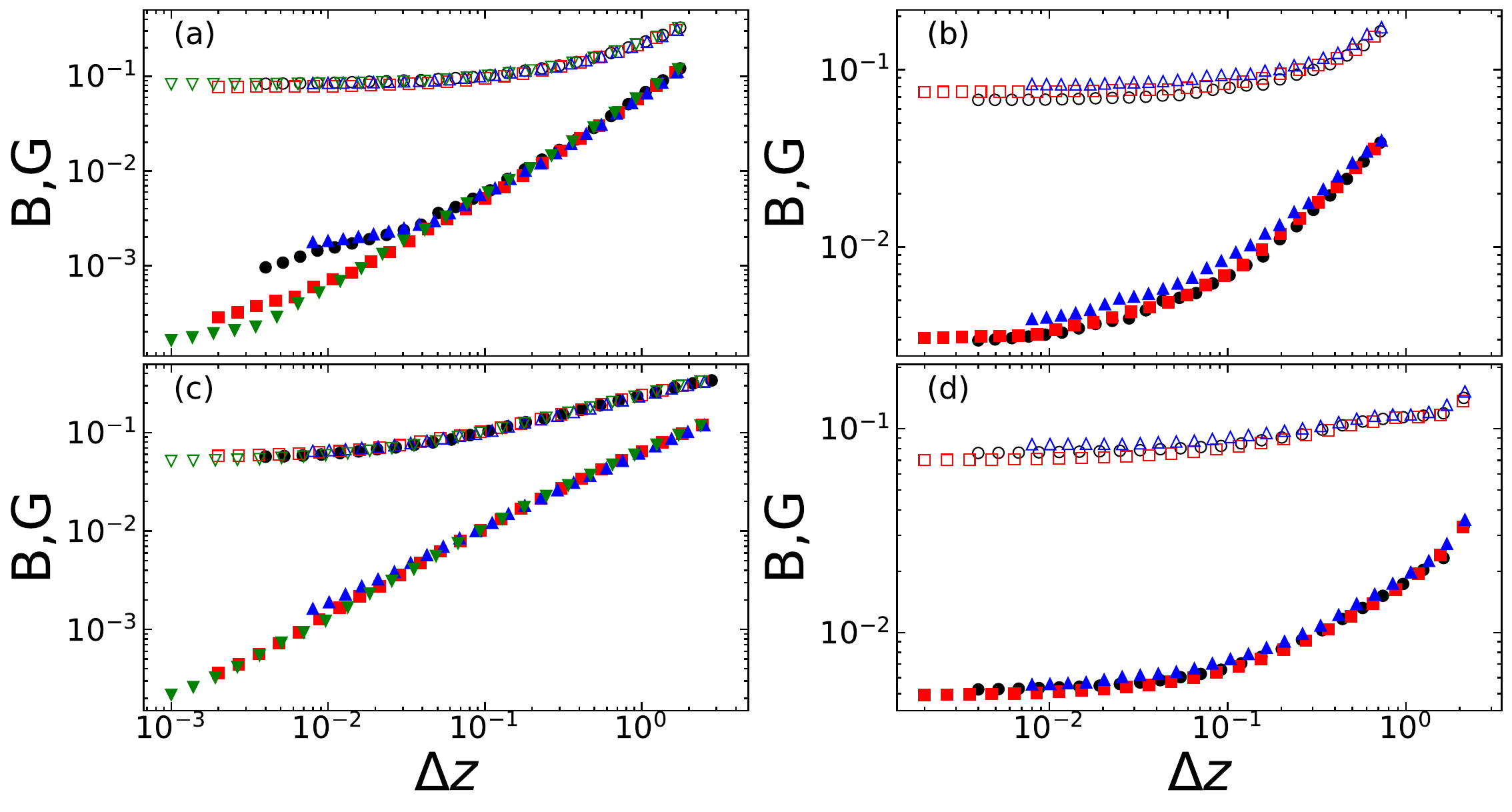}
    \caption{Bulk (open) and shear (filled) moduli of networks with $N$ = 256 (blue triangles), 512 (black circles), 1024 (red squares) or 2048 (green triangles) nodes for both random (a,b) and min $E_i$ (c,d) pruning as a function of $\Delta z$. The nominal prestrain used was 0.05. In (a,c) networks prepared under pure tension, while in (b,d) networks were prepared under pure shear. B approaches a non-zero constant in all panels, but G goes to 0 as $\Delta z \rightarrow 0$ in (a,c) and approaches a different constant in (b,d).}
    \label{fig:BG_finite}
\end{figure*}

For the bulk (B) and shear (G) moduli, finite size effects can obscure a finite discontinuity at $z = z_c$. We have ran simulations of both methods of pruning for both the tension and shear prestresses at a high nominal prestrain in order to determine if there is a finite discontinuity before loss of rigidity at small $z_c$. Fig.~\ref{fig:BG_finite} shows that for all types of prestress the bulk modulus shows a finite discontinuity, while the shear modulus only has a discontinuity for the system prestressed by pure shear.

\section{Scaling of shear modulus at fixed coordinations}

\begin{figure*}[ht]
    \centering
    \includegraphics[width=\textwidth]{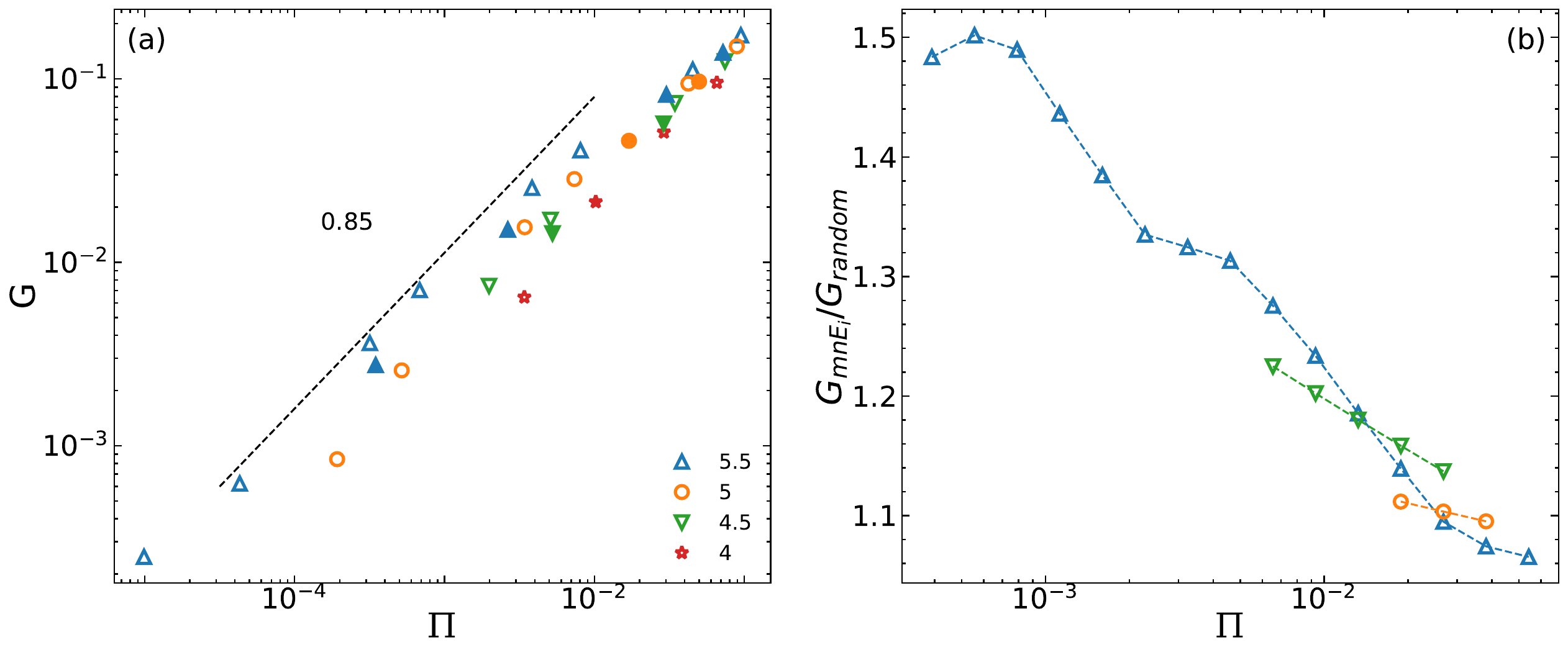}
    \caption{(a) Shear modulus of pruned networks at different values of $z$. For $z$ below the isostatic value, the shear modulus modulus depends on the tension on the network as a power law with exponent $0.85$ and this exponent is independent of pruning protocol. Open symbols are min $E_i$ pruned networks while solid symbols are randomly pruned. (b) At fixed $z$, the ratio of the modulus of $min E_i$ pruned networks to randomly pruned networks decreases and approaches 1 as tension increases. The slope decreases with decreasing $z$, and at sufficiently low $z$ the ratio is not bigger than $1.2$ for any tension.}
    \label{fig:SM_BG_fixed_z}
\end{figure*}

Fig. \ref{fig:SM_BG_fixed_z}(a) shows that for a fixed mean coordination $z$,networks display the same scaling of the shear modulus with tension, independent of $z$ or the pruning strategy. This exponent has already been reported in the literature, where the scaling was calculated by varying the tension applied to networks at a fixed $z$. The bulk modulus is relatively constant across the whole range of prestrain for both pruning methods. 

Fig. \ref{fig:SM_BG_fixed_z}(b) shows the ratio of the shear modulus of min $E_i$ pruned networks to the modulus of randomly pruned networks at the same values of $z$ as panel (a). At high tensions, this ratio approaches 1, implying that under very high tension the shear modulus is insensitive to pruning strategy.

\end{document}